\documentclass{llncs}

\usepackage{xspace,amsmath,url}
\usepackage{arydshln}

\usepackage{graphics}
\usepackage[outdir=./]{epstopdf}
\usepackage{epsfig}

\usepackage{grffile} 

\usepackage{caption}
\usepackage{subfigure}   

\newcommand{\exclude}[1]{}

\begin{document}

\title{A Bloom filter based semi-index on $q$-grams}

\author{Szymon Grabowski,
Robert Susik
and
Marcin Raniszewski
}

\institute{
	  Lodz University of Technology, Institute of Applied Computer Science,\\
	  Al.\ Politechniki 11, 90--924 {\L}\'od\'z, Poland
	  \email{\{sgrabow|rsusik|mranisz\}@kis.p.lodz.pl}
}

\maketitle

\begin{abstract}
We present a simple $q$-gram based semi-index, which allows to look for 
a pattern typically only in a small fraction of text blocks.
Several space-time tradeoffs are presented. 
Experiments on Pizza~\&~Chili datasets show that our solution 
is up to three orders of magnitude faster than the 
Claude et al.~\cite{CNPSTjda10} semi-index at a comparable space usage.
\end{abstract}

\section{Introduction}
\noindent 
String matching algorithms have traditionally been divided 
into on-line and off-line ones.
This distinction is not always satisfactory.
On-line solutions often lack in performance, if run over large 
texts, while off-line (index) schemes are complicated and/or resource-hungry.
In~\cite{CNPSTjda10} an intermediate approach was called a {\em semi-index}.
A semi-index is a data structure built on top of a given text, 
which is easy and fast to construct, uses a reasonable amount of memory 
(both during construction and its lifetime) and allows to search for a 
pattern faster than using an on-line scan (albeit typically not as fast 
as with a ``real'' index).

The semi-index of Claude et al.~\cite{CNPSTjda10} replaces the original text 
with a shorter one, namely such that some symbols of the alphabet are omitted.
The same symbols are also removed from the pattern before the search 
and potential matches have to be verified.
Clearly, the search speed is usually improved if a bigger part of the alphabet 
is sampled out, yet the problem is that in an extreme case the whole pattern 
may be ``erased''.  
Nonetheless, for long patterns ($m = 100$) a speedup in online search 
by factor about 5 while using 14\% extra space was reported. 
For moderately long patterns ($m = 20$) the speedup was less than twofold.
Several others algorithms from the literature can also be classified 
as semi-indexes, in particular $q$-gram based inverted files~\cite{PST2006}.

We use the standard notation throughout the paper. 
The pattern $P[0 \ldots m-1]$ is sought over the text $T[0 \ldots n-1]$.
Both strings are composed of symbols from a common integer alphabet 
$\Sigma = \{0, 1, \ldots, \sigma - 1\}$.

\section{Our algorithm}
\label{sec:alg1}

In this section we propose Bloom Filter based Semi-Index (BFSI), 
an algorithm combining highly selective filtering of text blocks 
before the actual search (text scan) 
with simplicity, 
both on the conceptual and implementational level.

The text $T$ is partitioned into $n/b$ fixed-size blocks of $b$ symbols, 
and successive blocks are grouped into superblocks of size $r$ blocks.
The overlapping $q$-grams of each $i$th superblock are added to a 
Bloom filter (BF)~\cite{Bloom1970}, 
represented as one bit table $\mathcal{B}_i$ of size $cbr$ bits, 
where $c$ is the chosen number of bits per item in a BF, 
trading its size for accuracy.
Let us introduce a set of $u$ baseline hash functions, 
$h_k(S): \sigma^q \to \{0, 1, \ldots, cbr-1\}$, 
for $k \in \{0, \ldots, u-1\}$, where $S$ is a string of length $q$.
The actual hash functions applied to elements from block 
$ir+j$, for any $0 \leq i \leq (n/(br))$ and $0 \leq j < r$, 
are however of the form $h'_k(S) = h_k(S) \cdot r + j$.
Note that the hashes for $q$-grams from $j$th block within $i$th superblock 
may affect only such $\mathcal{B}_i[\ell]$ cells that $\ell\ \text{mod}\ r = j$.
%
In other words, within a superblock the $q$-grams for each block 
are as if stored in a separate BF, 
but these $r$ substructures are interwoven; 
the motivation for choosing such a layout will be given later.

\begin{figure}
\centerline{
\includegraphics[width=0.99\textwidth,scale=1.0]{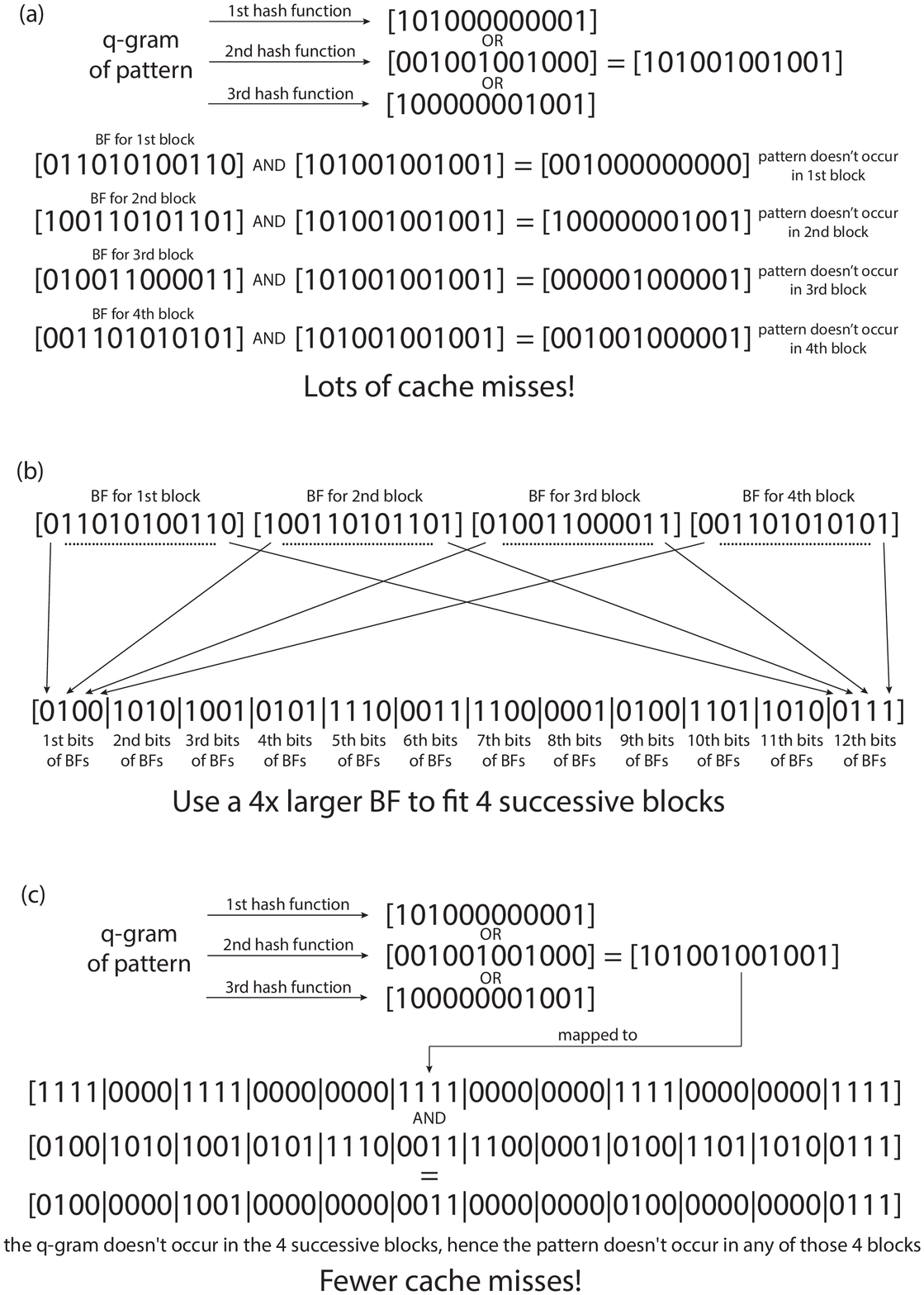}
}
\caption[Illustration]
{Searching of a pattern $q$-gram in a Bloom filter (BF) in two variants: 
(a) standard searching in each BF block causes a lot of cache misses, 
(c) searching in a larger BF structure created by interwoved BF blocks, 
which (b) reduces the number of cache misses thanks to the changed data layout}
\label{fig:semi_all}
\end{figure}

The search idea for the pattern $P[0 \ldots m-1]$, where $m \geq q$, 
is very simple;
if we cannot exclude that all its $q$-grams occur (in any location) 
in a given block, then the block is scanned using some ``off-the-shelf'' 
exact string matching algorithm.
In the opposite case, when we are sure that at least one $q$-gram of $P$ 
has no occurrence in the block, the block is skipped.
To this end, the search starts with computing $u$ hash values
for each of the $(m-q+1)$ overlapping $q$-grams of $P$.
To simplify the exposition, let us present the following procedure 
on a single $q$-gram of the pattern, e.g. the first one, $P[0 \ldots q-1]$.
Its computed hashes are of the form $h_k(P[0 \ldots q-1])$, for 
$k \in \{0, \ldots, u-1\}$.
We traverse over successive superblocks, and in each superblock
we need to check in which of the $r$ blocks the $q$-gram occurs.
To this end, we calculate $h'_k(P[0 \ldots q-1]) = h_k(P[0 \ldots q-1]) \cdot r + j$, 
for all valid $k$ and $j$, and the set bits in the found positions denote 
the blocks containing the current $q$-gram.
If a given block contains a set bit for all the $q$-grams from the pattern, 
we have to scan the block for occurrences of $P$ 
(as the actual pattern matching algorithm for it, we chose FAOSO~\cite{FGjda09}).
If not, we proceed to the next block.
Yet, for the next block the accessed cells of $\mathcal{B}_i$ are simply 
successors of the corresponding cells accessed in the previous block 
(since we replace term $j$ with $j+1$).
This contiguous access pattern is cache-friendly, hence its justifies our 
data layout in $\mathcal{B}_i$.
Fig.~\ref{fig:semi_all} illustrates.
%
%

Longer $q$-grams should be more selective than shorter ones, 
yet choosing a too large $q$ prevents searching for short patterns.
Another observation is that we may trade the BFSI space for filtering 
selectivity with building the BF with a sample of $q$-grams only.
To this end, we consider the following variants:
\begin{itemize}
\item STD, the standard version of BFSI, with all $q$-grams from the block used,
\item SAM (sampled $q$-grams), which inserts every $s$th, $s \geq 1$, 
sampled $q$-gram to a BF; note that $s=1$ corresponds to the STD variant,
\item MSAM (minimizer-based sampled $q$-grams), which is similar to SAM, 
but samples the $q$-grams in a non-regular way, using the idea of 
minimizers~\cite{RHHMY2004}.
More concretely, we slide a window of size $w$ over a block 
and in each window we find the lexicographically smallest substring of length $p$, 
where $p < w$.
Note that successive text windows are likely to share the same minimizer.
The starting symbols of all distinct minimizers in a block are also the starting 
positions of the sampled $q$-grams.
It is easy to notice that the gap between two successive sampled $q$-grams 
cannot exceed $w - p + 1$.
\end{itemize}

\section{On using $q$-grams for string matching}

The notion of $q$-grams 
is widely used in string matching algorithms for more than two decades.
In 1992 Ukkonen~\cite{Ukk1992} introduced the $q$-gram distance between 
two strings and used it for (online) approximate string matching.
Takaoka~\cite{T1994} presented an approximate matching filter based on 
sampling $q$-grams from the text.
This technique was refined by Sutinen and Tarhio~\cite{ST1995} 
with using ordered $q$-grams, based on a simple observation that 
the preserved $q$-grams must be approximately at the same locations 
both in the pattern and its approximate match in the text.
Another application of $q$-grams (also ordered ones) for efficient 
approximate pattern matching was given by Fredriksson and Navarro~\cite{FNjea2004}, 
in an approach which can be classified as a member of the Boyer--Moore family.
Burkhardt and K{\"a}rkk{\"a}inen~\cite{BK2001} advocated for gapped $q$-grams, 
proving their superior filtering capabilities.
Fredriksson and Grabowski~\cite{FG2006} applied a byte code over 
successive (non-overlapping) 
$q$-grams of the text, to allow compressed pattern matching over 
arbitrary texts with a simple application of virtually any multiple pattern 
matcher (if only the pattern length is at least $2q-1$; 
shorter patterns are handled with a different, slower, algorithm).

Indexes on $q$-grams have not once been applied for searching 
biological sequence databases.
In particular, QUASAR~\cite{BCFLRV1999} and Swift~\cite{RSM2006} basically 
divide the text into small blocks and only the blocks having at least 
a specified fraction of $q$-grams in common with the query sequence 
are processed carefully, e.g., with BLAST, to report alignments.
In some solutions, e.g., BLAT~\cite{Kent2002blat}, the index contains 
only the information about the non-overlapping $q$-grams, resulting 
in a reduction in the index size but a loss in sensitivity.
Although our solution presented here, with all or sampled $q$-grams, 
may resemble the listed $q$-gram based indexes for biological data, 
we are not aware of using a Bloom filter (or similar succinct and lossy membership 
data structure) for storing the $q$-grams.


\section{Experimental results}
\label{sec:exp}

In order to evaluate the performance of BFSI, we run quite extensive 
experiments.
As the competitor we took the algorithm from Claude et al., whose 
source codes were received from the authors.
The test machine 
was equipped with an Intel Core i3-2100 CPU clocked at 3.1\,GHz,
4\,GB of DDR3-RAM (1333\,MHz), 
running Debian 3.2.63 x86\_64. 
Our algorithm 
was implemented in C++ 
and compiled with \texttt{g++} 4.8.1 with \texttt{-O3}. 
As the datasets we took 
three 
50\,MB texts from the 
widely used Pizza~\&~Chili corpus (\url{http://pizzachili.dcc.uchile.cl/}).
For each test case, 
we search 100 random patterns 
and present the average 
timings.
BFSI tests were ran with varying multiple parameters.
We can distinguish 
parameters common for all variants (STD, SAM and MSAM), 
parameters specific for particular variants (SAM and MSAM) 
and parameters of the chosen search algorithm 
(which was FAOSO in all tests). 
The standard variant, STD, uses only the common parameters:
$q$-gram size ($q$), block size ($b$) and Bloom filter density 
($c$), which affects the number of hash functions and the expected 
false positives rate. 
%
The parameter specific to SAM is $s$, 
which is the 
$q$-gram sampling rate 
(i.e., every $s$-th $q$-gram is selected).
The MSAM variant makes use of two specific parameters: 
window size ($w$) and minimizer length ($p$). 

We set the necessary requirements:
$m \geq q + s - 1$ for the SAM variant 
and
$m \geq \max(q, w, w + q - p)$ for the MSAM variant.
As said above, in all variants we use the FAOSO exact pattern matching 
to scan the selected text blocks.
FAOSO depends on two parameters which were fixed 
($U = 4$ and $k = 2$). 
Note it might be possible to achieve better performance with tuning the FAOSO 
parameters for particular datasets or use another pattern matcher instead.


\begin{figure}[pt]
\centerline{
\includegraphics[width=0.49\textwidth,scale=1.0]{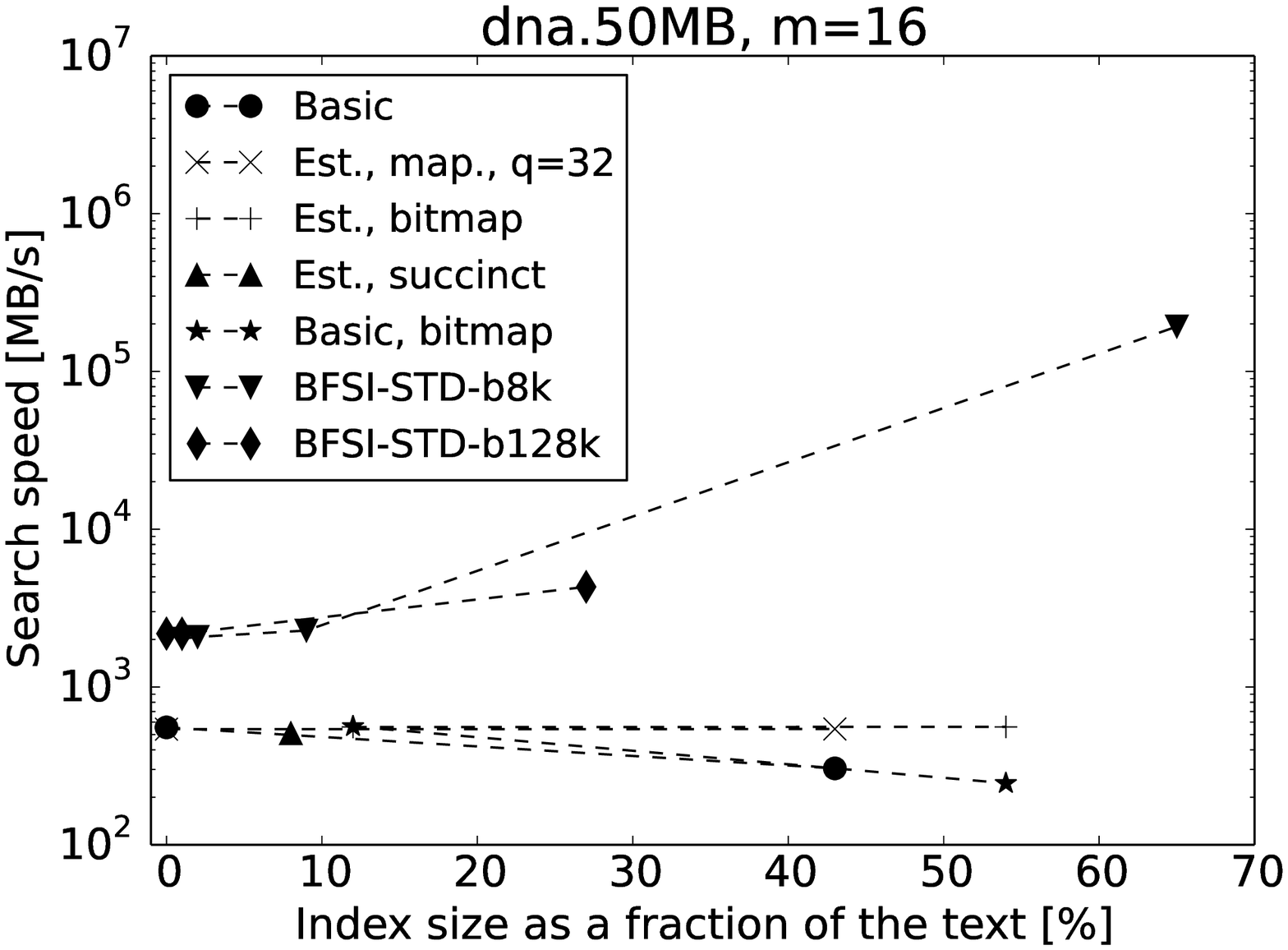}
\includegraphics[width=0.49\textwidth,scale=1.0]{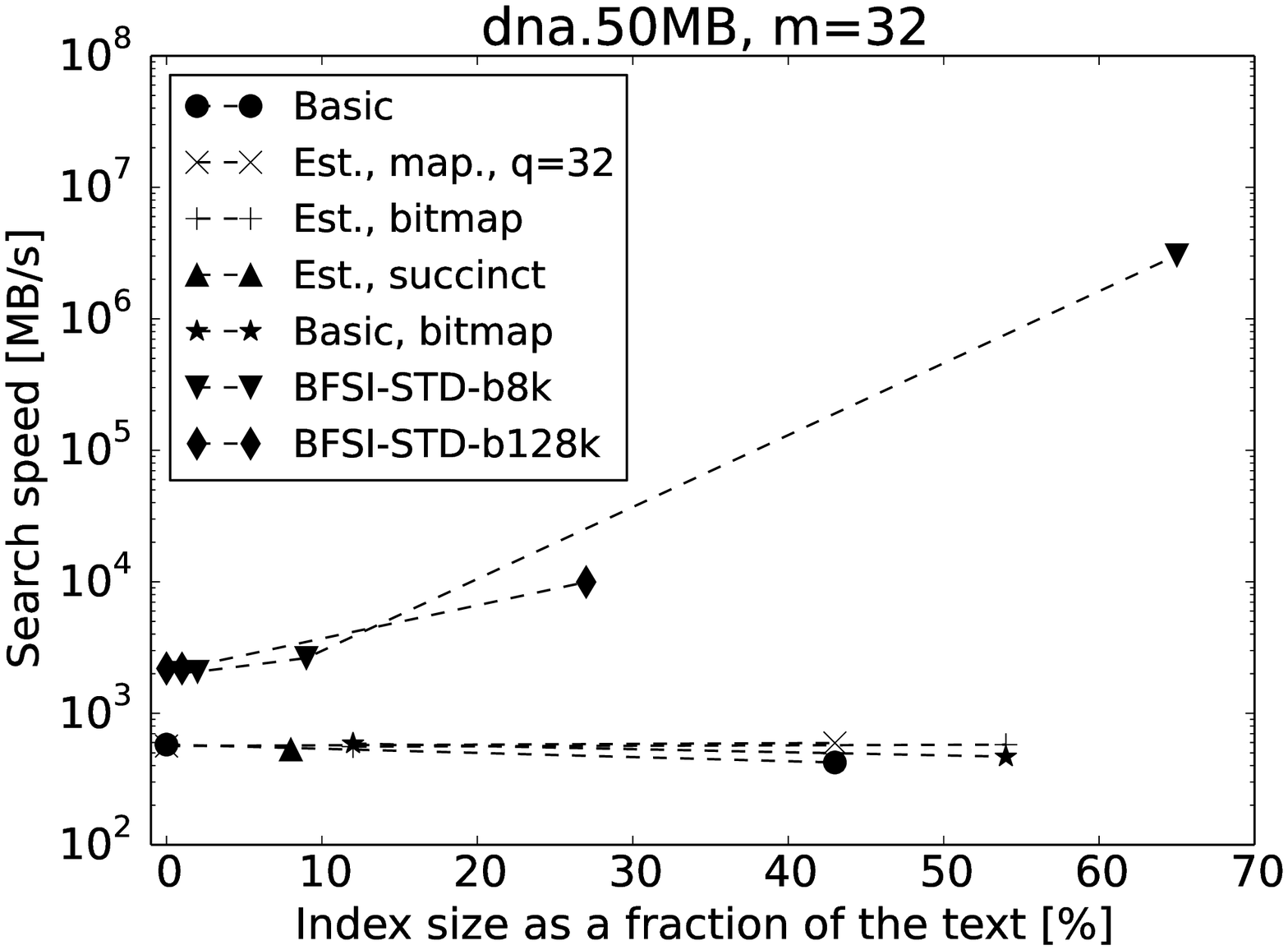}
}
\centerline{
\includegraphics[width=0.49\textwidth,scale=1.0]{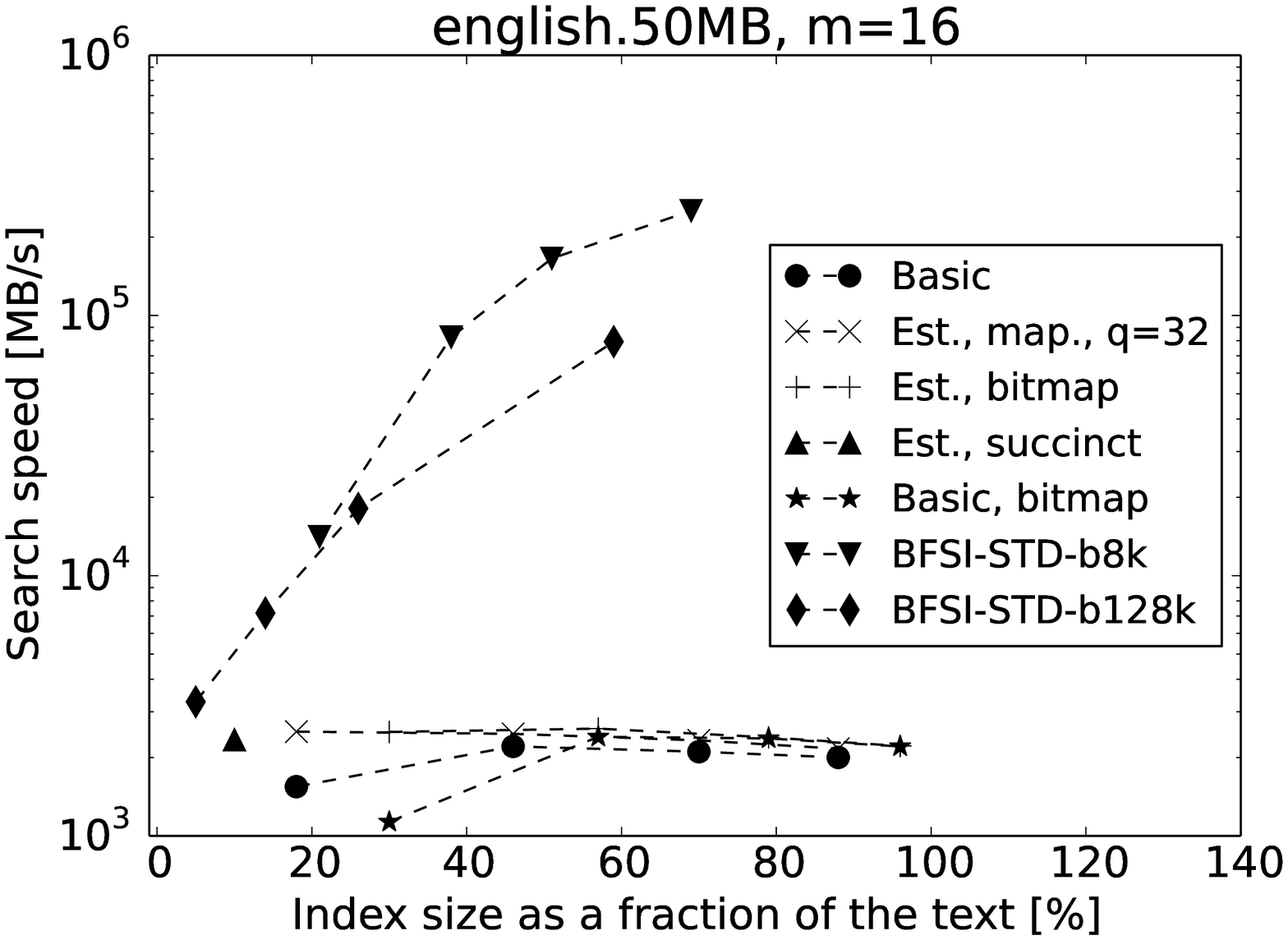}
\includegraphics[width=0.49\textwidth,scale=1.0]{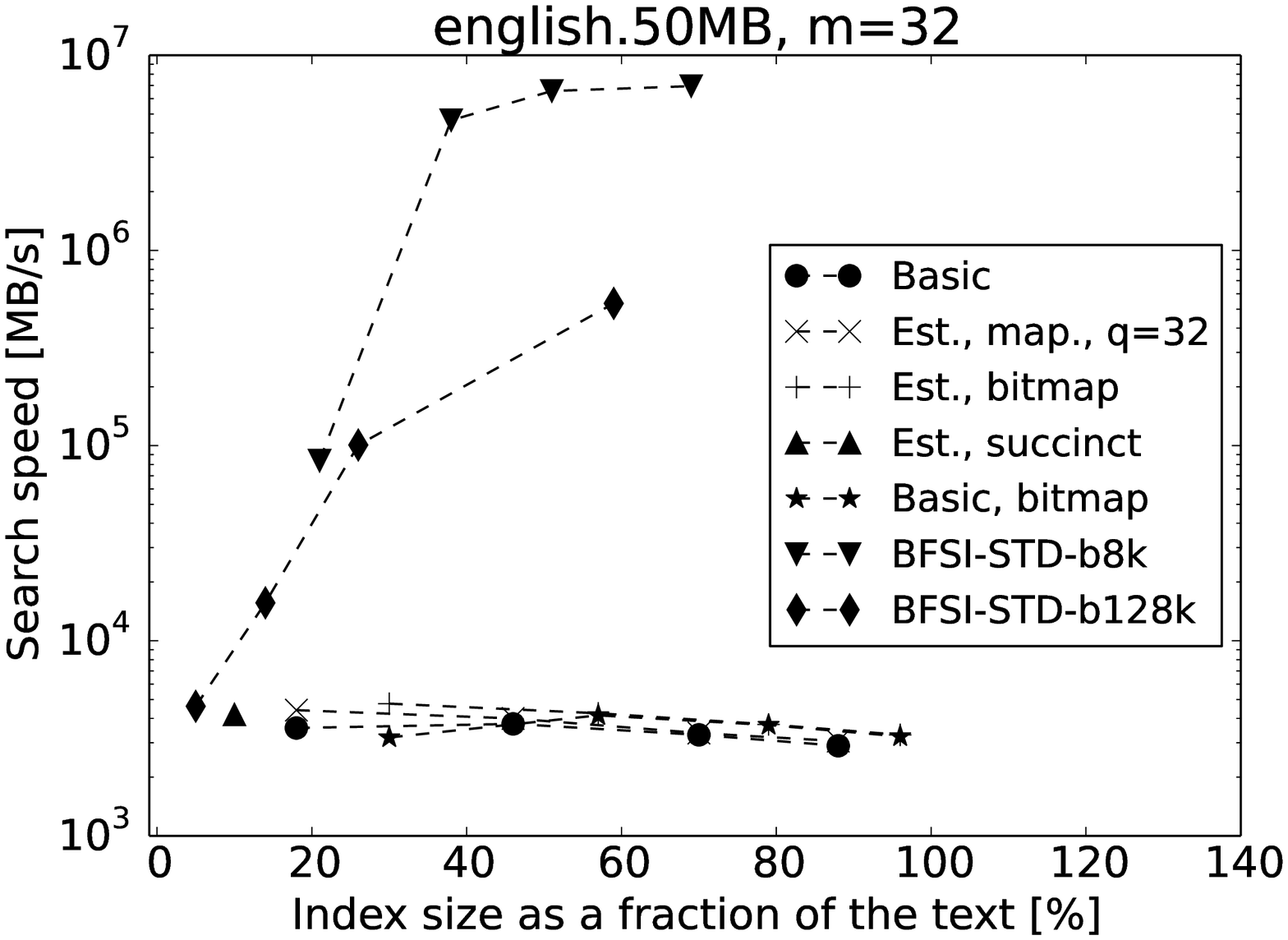}
}
\caption[Results]
{Speed/Space efficiency comparison of Succinct semi-index to BFSI}
\label{fig:speed_space_perf}
\end{figure}

Fig.~\ref{fig:speed_space_perf} presents a comparison of BFSI 
and Succinct semi-index~\cite[Sect.~2.2 and Sect.~3.1]{CNPSTjda10}; 
see the information on the particular variants in the cited work. 
Here we used the STD variant of BFSI with the parameter values of 
$q \in \{3,4,5,8\}$, $c = 6$ and $b \in \{8, 128\}$ (kilobytes). 
We show the space used by the index (in addition to the text itself), 
as a fraction of the text size, and the search speed in MB/s.
Our solution 
in most cases is about three orders of magnitude faster 
than Succinct semi-index at
the same index size. 
As expected, choosing a larger $q$ makes the search faster yet for the 
price of requiring more space.


\begin{figure}[pt]
\centerline{
\includegraphics[width=0.49\textwidth,scale=1.0]{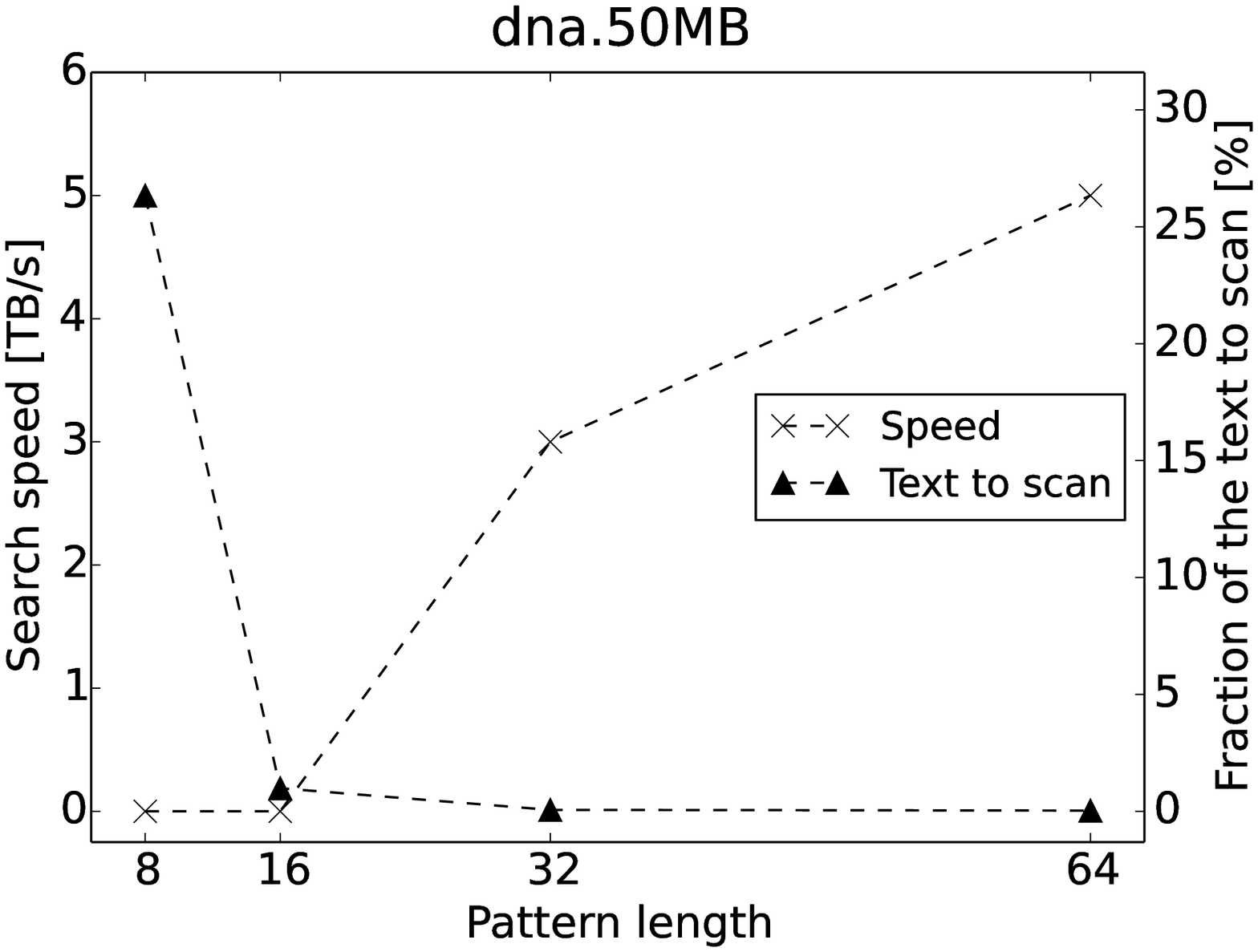}
\includegraphics[width=0.49\textwidth,scale=1.0]{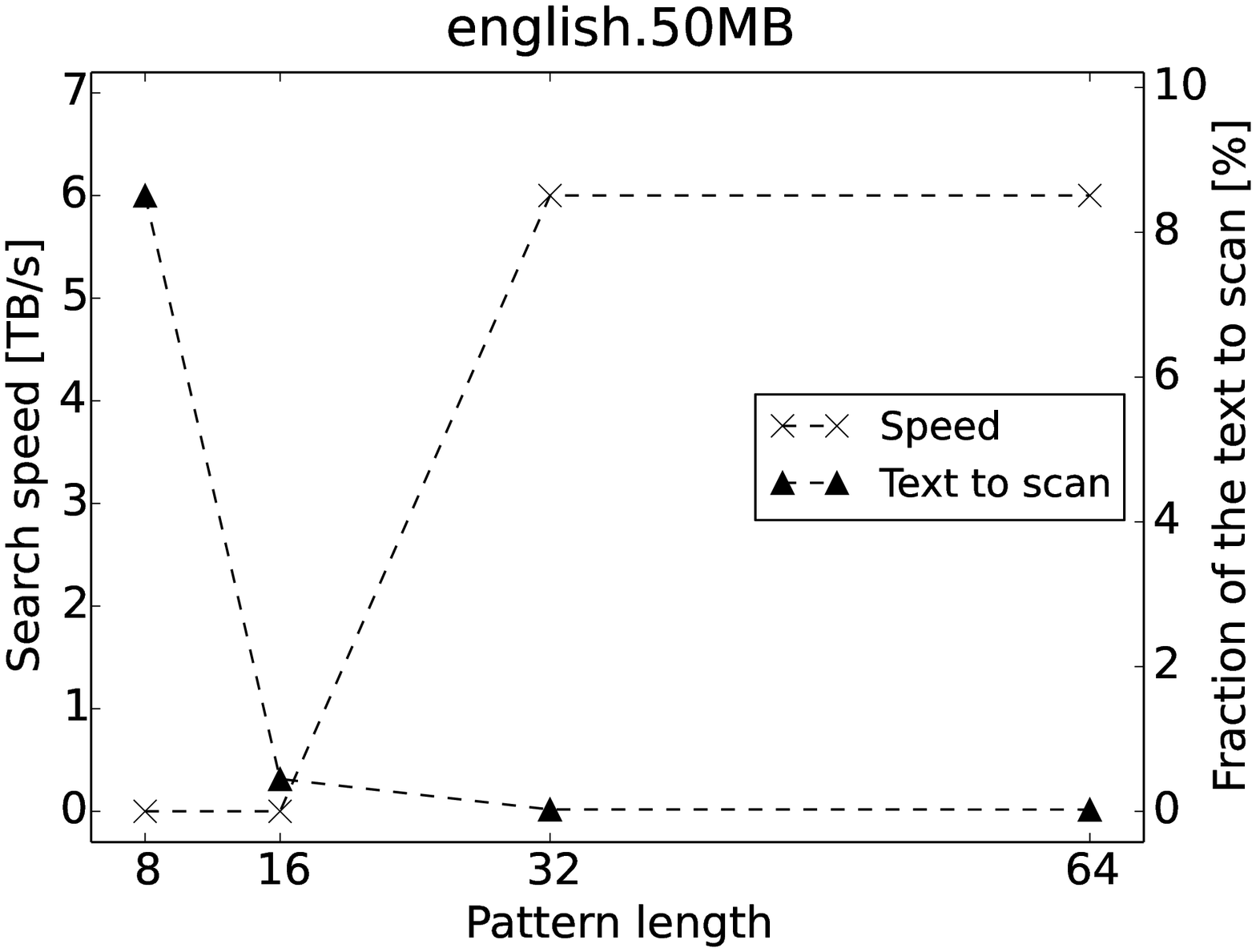}
}
\caption[Results]
{Search speed of BFSI-STD variant for varying pattern length $m$. Fixed parameters: $b=8$ (KB), $c=6$, $q=8$.}
\label{fig:speed_space_m_threeY}
\end{figure}

Fig.~\ref{fig:speed_space_m_threeY} presents performance comparison 
in function of pattern length $m$. 
The plot shows (especially in the case of \texttt{dna}) that speed 
grows more or less in a linear manner with increasing $m$.
On one hand, more $q$-grams in a longer pattern translates to rapidly 
decreasing number of blocks to scan (as the axis ``Fraction of the text 
to scan'' shows), but on the other hand the $q$-gram checks in the Bloom filters 
are not free and will eventually be dominating.
Yet another factor 
is faster 
search for longer 
pattern (a feature common to most fast pattern matchers, including FAOSO).


\begin{figure}[pt]
\centerline{
\includegraphics[width=0.49\textwidth,scale=1.0]{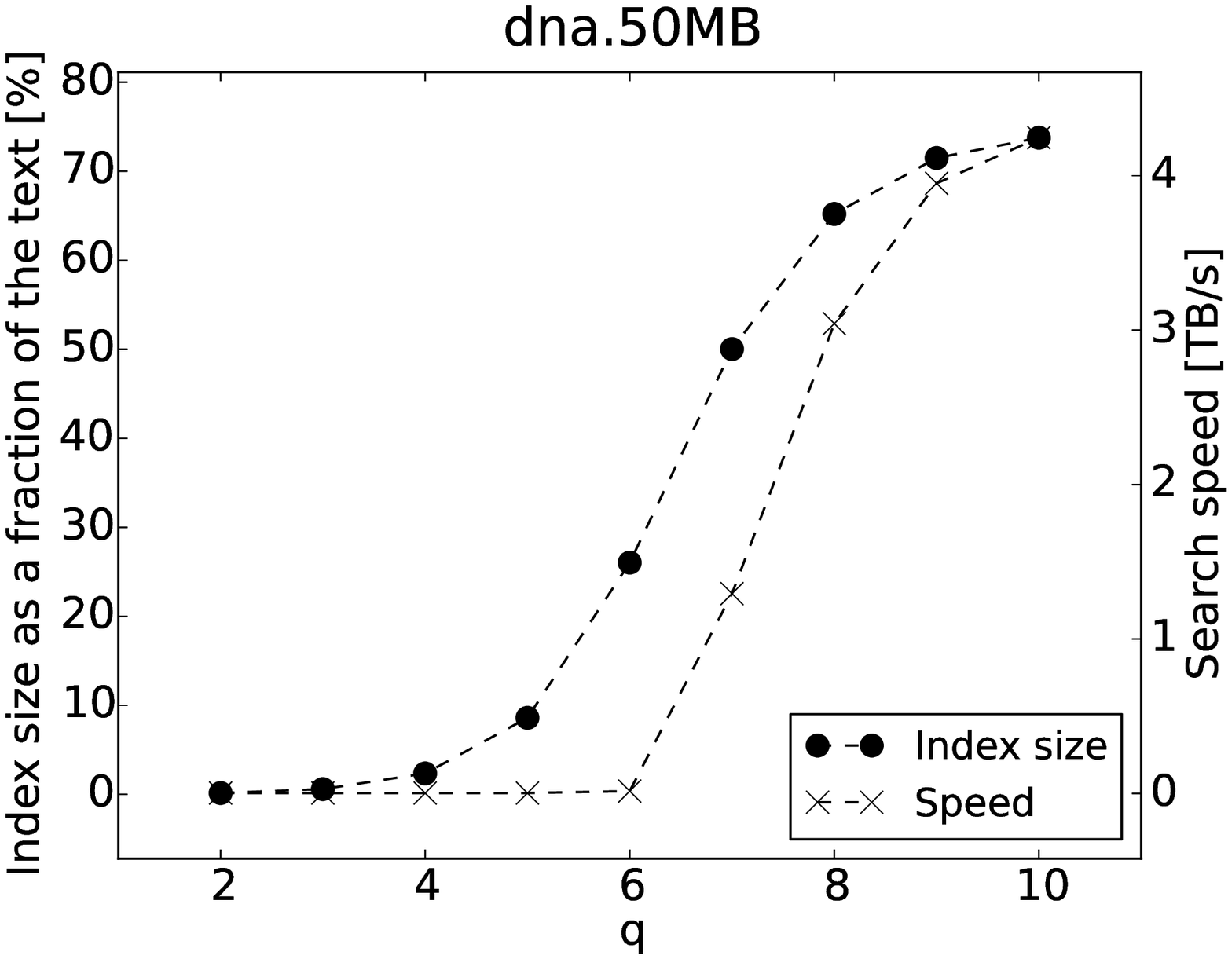}
\includegraphics[width=0.49\textwidth,scale=1.0]{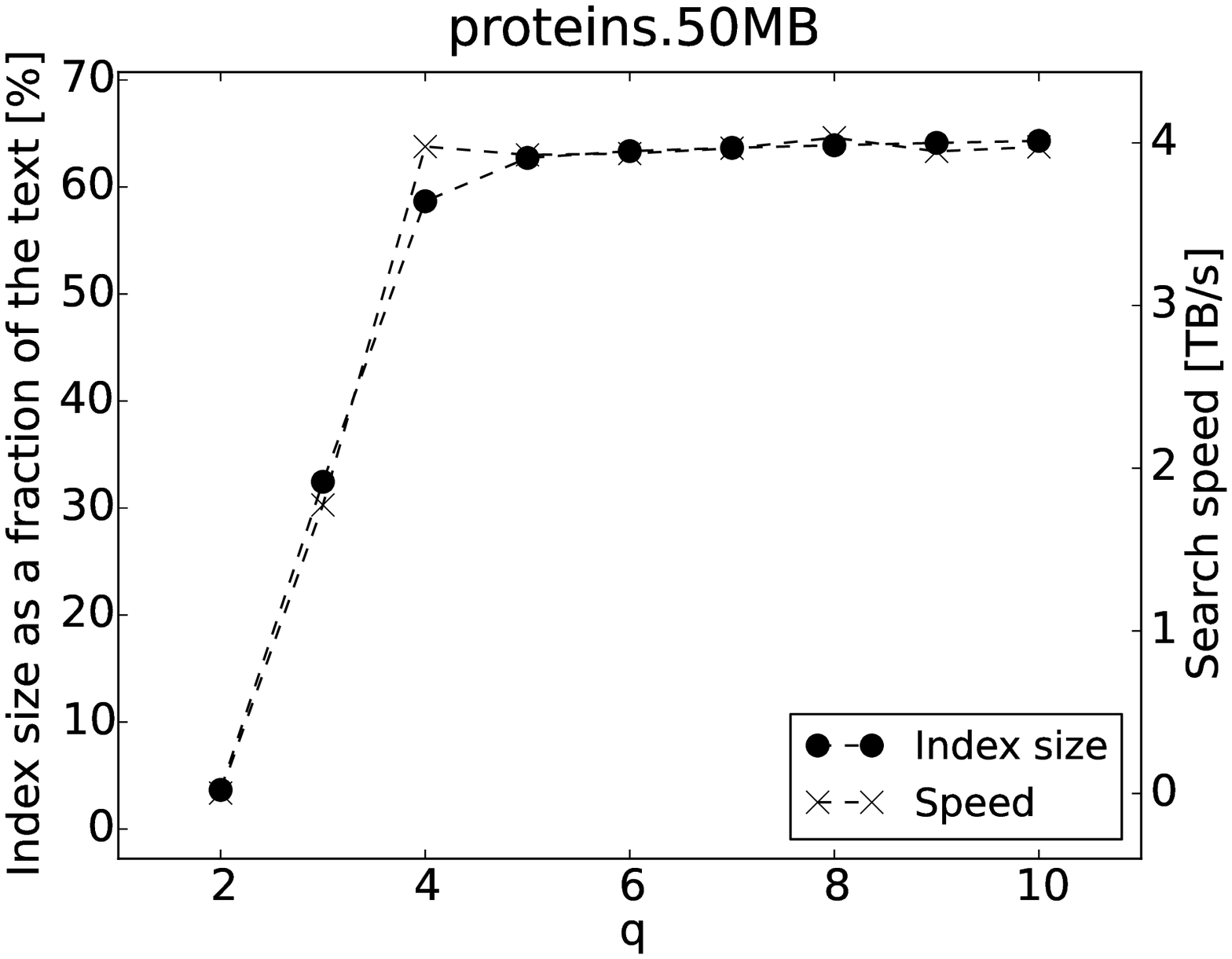}
}
\caption[Results]
{Search speed and index size of BFSI-STD variant for varying $q$ and fixed 
$m=32$, $c=6$, $b=8$ (KB)}
\label{fig:speed_space_q_twoY}
\end{figure}

\begin{figure}[pt]
\centerline{
\includegraphics[width=0.49\textwidth,scale=1.0]{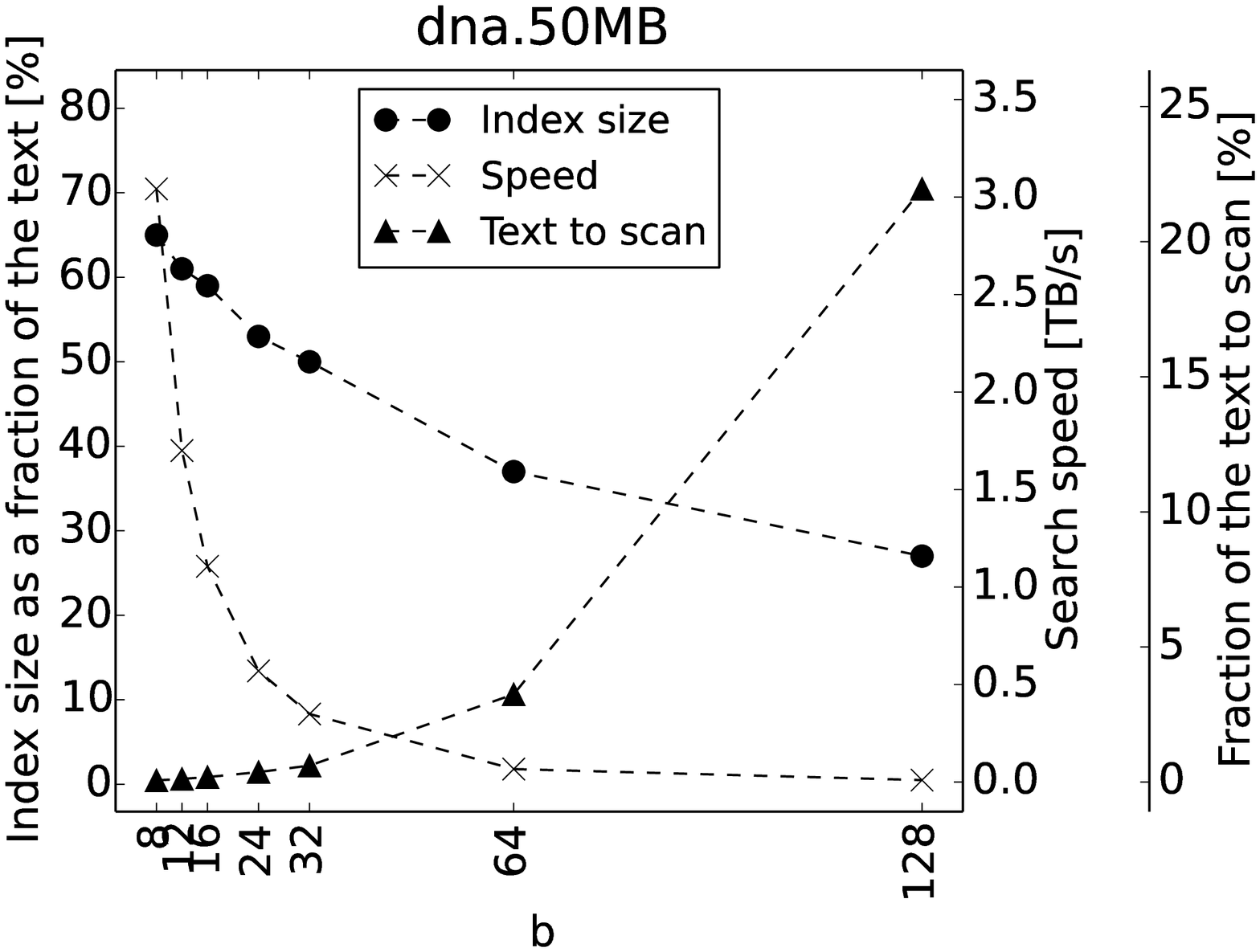}
\includegraphics[width=0.49\textwidth,scale=1.0]{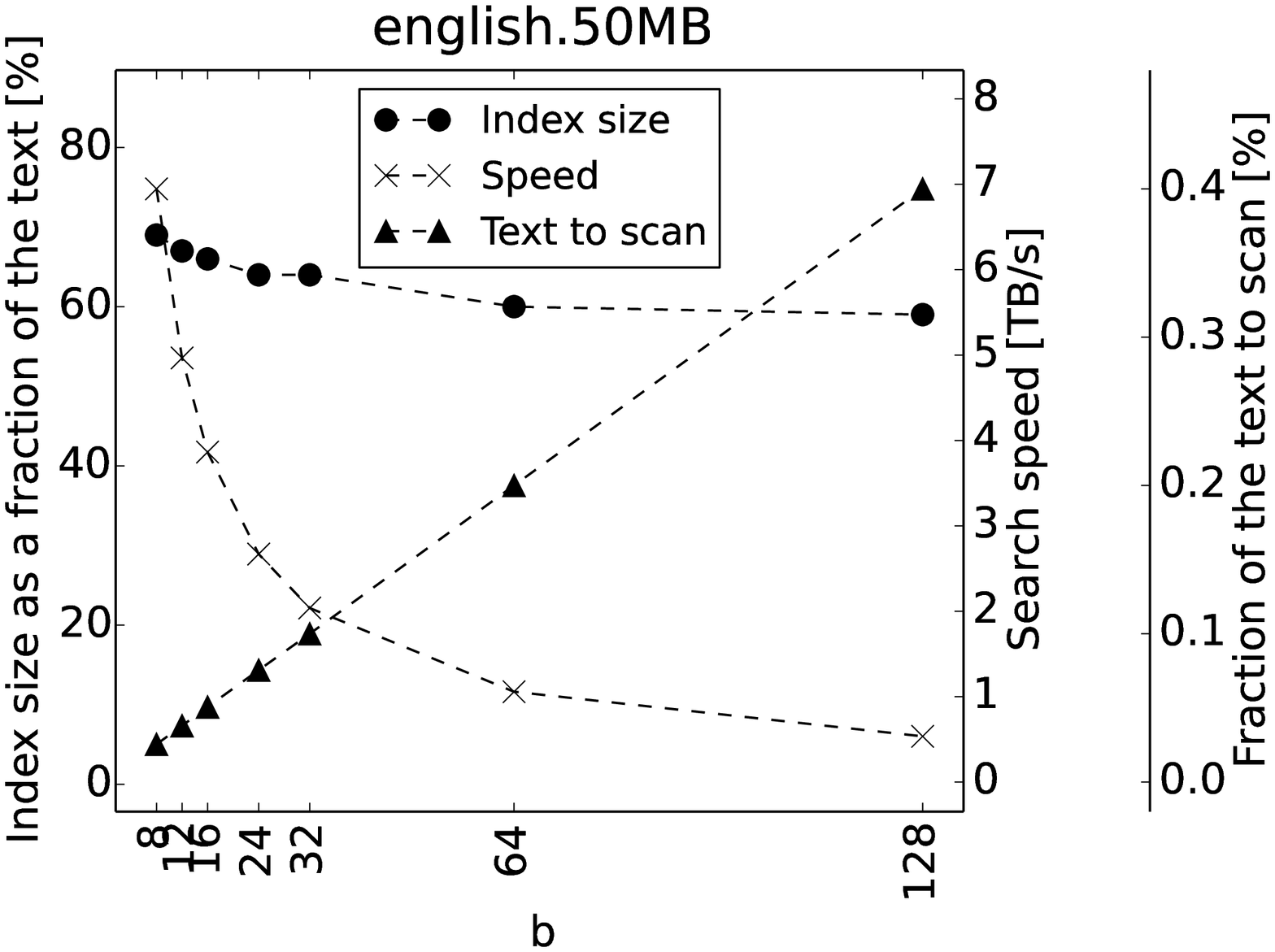}
}
\caption[Results]
{Search speed, index size and amount of text scanned by BFSI-STD for varying block size $b$ (in KB) and fixed $m=32$, $c=6$ and $q=8$}
\label{fig:speed_space_bs_threeY}
\end{figure}

Next we examine how varying $q$, from 2 to 10, affects 
the search speed and the index space (Fig.~\ref{fig:speed_space_q_twoY}).
Roughly speaking, $q$ is the most important parameter for BFSI performance.
For a smaller alphabet (\texttt{dna}) a larger $q$ is required to achieve 
high block selectivity and thus also the overall speed, while for a larger alphabet (\texttt{proteins})
the ``saturation'' is achieved earlier.
Fig.~\ref{fig:speed_space_bs_threeY} presents speed and space in function of block size $b$.
As expected, with increasing block size, the index gets smaller and the performance decreases. 
The drop in speed is however much higher than the saving in index space 
(especially for small values of $b$).


\begin{figure}[pt]
\centerline{
\includegraphics[width=0.49\textwidth,scale=1.0]{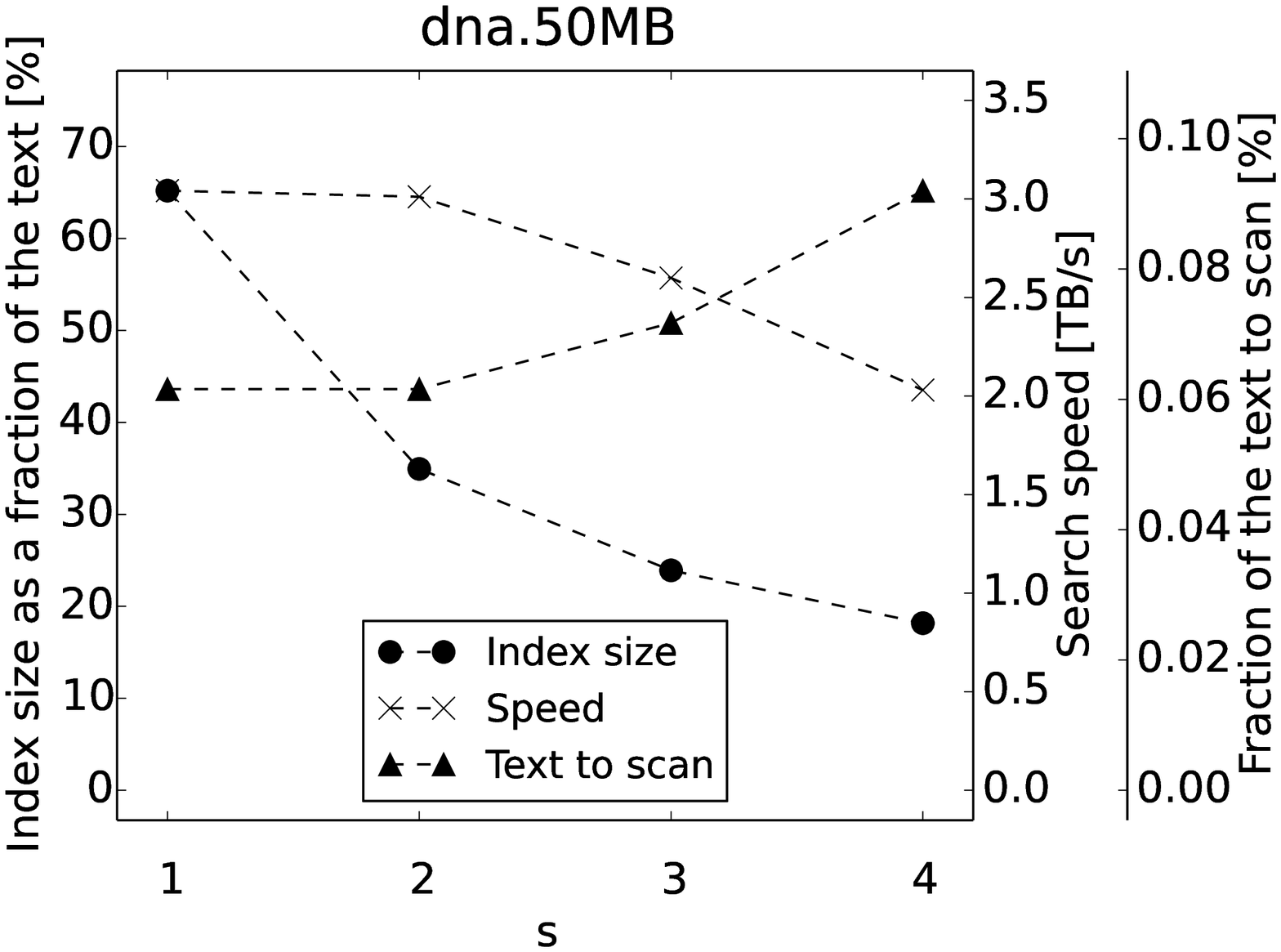}
\includegraphics[width=0.49\textwidth,scale=1.0]{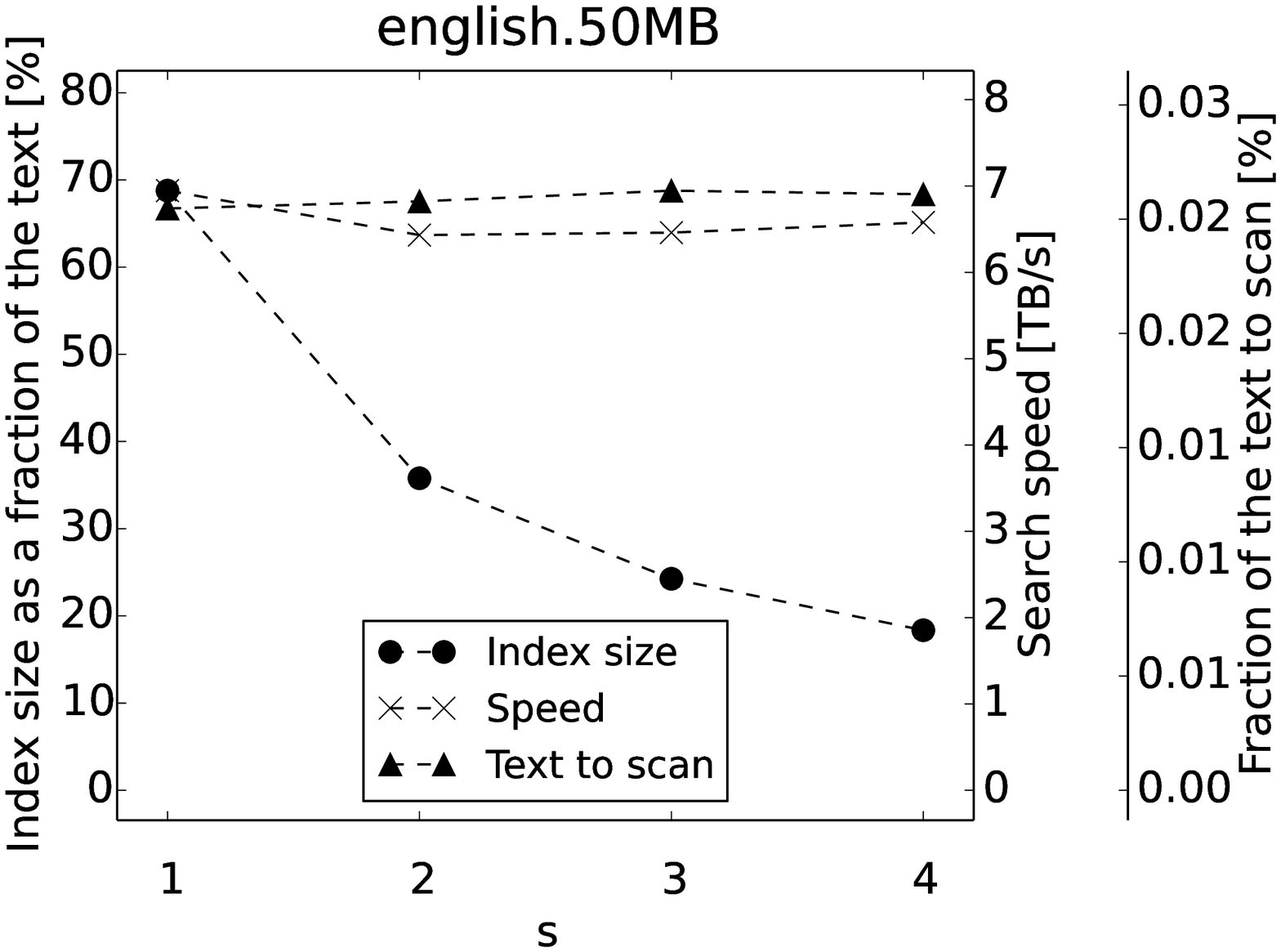}
}
\caption[Results]
{Search speed, index size and the amount of text scanned by BFSI-SAM for varying $s$ and fixed $m=32$, $c=6$, $b=8$ (KB), $q=8$}
\label{fig:speed_space_g_threeY}
\end{figure}




Fig.~\ref{fig:speed_space_g_threeY} shows the impact of the sampling 
parameter $s$ on the search speed, index size and the fraction of text to scan 
in BFSI-SAM.
Larger $s$ can significantly reduce the index size,
but the performance is reduced as well (although not much for \texttt{english} and relatively long patterns, 
as seen on the right figure). 
%

\begin{figure}[h]
\centerline{
\includegraphics[width=0.54\textwidth,scale=1.0]{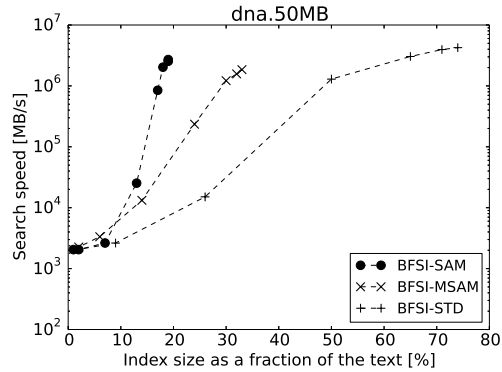}
}
\caption[Results]
{Speed/space relation for fixed $m=32$, $c=6$, $s=4$, $p=4$, $w=4$, $b=8$ (KB) and $q \in \{3,\ldots,10\}$}
\label{fig:fig_dna_speed_our_m32_eps}
\end{figure}

Fig.~\ref{fig:fig_dna_speed_our_m32_eps} presents the performance of our algorithms in function of index size.
BFSI-MSAM is slightly better when index size is below 9\%. 
For a bigger index the speed of BFSI-SAM is drastically increasing. 
Despite index size, both variants have similar performance. 
However, BFSI-STD reaches the highest speed 
(above 4\,TB/s) 
if significant space for the index is allowed.



\section*{Acknowledgement}
The work was supported by the Polish National Science Centre under the project DEC-2013/09/B/ST6/03117 (the first and the third author).

\bibliographystyle{abbrv}
\bibliography{semi}

\end{document}